\documentclass[oldversion]{aa}
\usepackage{graphics,epsf,epsfig,graphicx}

\begin{document}

\title{Keck spectroscopy and Spitzer Space Telescope analysis of the
  outer disk of the Triangulum
  Spiral Galaxy M33}

\author{David L. Block \inst{1,4}, Fran\c coise Combes \inst{2}, 
Iv\^anio Puerari \inst{3},
Kenneth C. Freeman \inst{4}, Alan Stockton\inst{5}, Gabriela Canalizo\inst{6}, 
Thomas H. Jarrett \inst{7}, Robert Groess \inst{1}, Guy Worthey
\inst{8}, 
Robert D. Gehrz \inst{9},
Charles E. Woodward \inst{9}, Elisha F. Polomski \inst{9} and Giovanni G. 
Fazio \inst{10} }

\offprints{D.L. Block \email{block@cam.wits.ac.za}}
\institute{
School of Computational and Applied Mathematics,
University of Witwatersrand, Private Bag 3, WITS 2050, South Africa
\and
Observatoire de Paris, LERMA, 61 Av. de
l'Observatoire, F-75014, Paris, France
\and
Instituto Nacional de Astrof\'\i sica,
Optica y Electr\'onica,
Calle Luis Enrique Erro 1, 72840 Tonantzintla, Puebla, M\'exico
\and
Mount Stromlo and Siding Spring
Observatories, Research School of Astronomy and Astrophysics, Australian
National University, Australia
\and
Institute for Astronomy, University of Hawaii, 2680 Woodlawn Drive,
Honolulu, Hawaii, USA
\and
Institute of Geophysics and Planetary Physics and Department of Physics, University of California,
Riverside, CA 92521, USA
\and
Infrared Processing and Analysis Center,
100-22, CALTECH, 770 South Wilson Ave, Pasadena, CA 91125, USA
\and
Washington State University, 1245 Webster Hall, Pullman, WA
99163-2814, USA
\and
Department of Astronomy, University of Minnesota, 116 Church St. SE,
Minneapolis MN 55455, USA
\and
Harvard-Smithsonian Center for Astrophysics, 60 Garden St., Cambridge,
MA 02138, USA}
\date{Received XX XX, 2006; accepted xx xx, 2007}
\authorrunning{Block et al.}
\titlerunning{The Outer Disk of M33}

\abstract{In an earlier study of the spiral galaxy M33, we photometrically 
identified arcs or outer spiral arms of intermediate
age (0.6 Gyr $-$ 2 Gyr) carbon stars precisely at the commencement of the
HI-warp. Stars in the arcs were unresolved,   
but were likely thermally-pulsing asymptotic giant branch carbon stars.
Here we present Keck I spectroscopy of seven intrinsically bright and red 
target stars in the outer, northern arc in M33. The target stars 
have estimated visual magnitudes as faint as V$\sim$ 25$^{m}$. 
Absorption bands of CN are seen in all seven spectra
reported here, confirming their carbon star status. In addition, we
present Keck II spectra of a small area 0.5 degree away from the centre of
M33; the target stars there are also identified as carbon
stars. We also study the non-stellar PAH dust morphology of M33
secured using IRAC on board the Spitzer Space Telescope.
The Spitzer 8$\mu m$ image attests to a change of spiral phase at the start of
the HI warp. The Keck spectra confirm that carbon stars may safely be
identified on the basis of their red J$-$K$_{\rm s}$ colours in the outer, low
metallicity disk of M33. We propose that 
the enhanced number of carbon stars in the outer arms are an indicator
of recent star formation, fueled by gas accretion from the HI-warp 
reservoir.
}

\maketitle

\section{Introduction}

Like the two dominant spiral galaxies in the Local Group
(MW and M31), M33 is known to have a prominent warp.
This warp is spectacular in the HI-21cm line:
at many points the line of sight intersects the disk twice. In M33 
the warping commences at a radius of $\sim$ 5 kpc and at a
radius of $\sim$ 10 kpc the rotation axis of neutral hydrogen gas is
inclined by some 40 degrees to the axis of the inner disk,
as revealed by the tilted ring model of Rogstad
et al. (\cite{rog76}). More recently, 
Corbelli and Schneider (\cite{corbellischneider97})  confirm that at a
radius of 5 kpc (about 3 disk scalelengths), the HI distribution 
in M33 shows a distinct change in inclination.

Warps may be tidally generated when companions are obviously present;
but in galaxies such as M33 which do not have any close companions, 
warping and gas infall may be inextricably linked (for a review, see 
section 7 in Binney \cite{binney92}). 

In this paper, we focus our attention on the outer warped
disk of M33, and in particular on its red stellar population.
If the HI-warp in M33 is induced by the infall of gas, then one 
interesting  confirmation would be the presence of a red intermediate age  
(0.6 Gyr - 2 Gyr) population of carbon stars at the locale of the HI-warp. 
 Indeed, C-stars are expected to be associated with recent star formation.
Tsalmantza et al. (2006) have found them towards the center
in the LMC and associated with spiral arms in M31, the loci of
star formation. In the SMC and in M33, their sample size was too small 
to trace their radial distribution, but Rowe et al. (2005)
found them also relatively more abundant in the outer parts of M33.

\begin{figure*}[ht]
\vspace{17.0cm}
\includegraphics{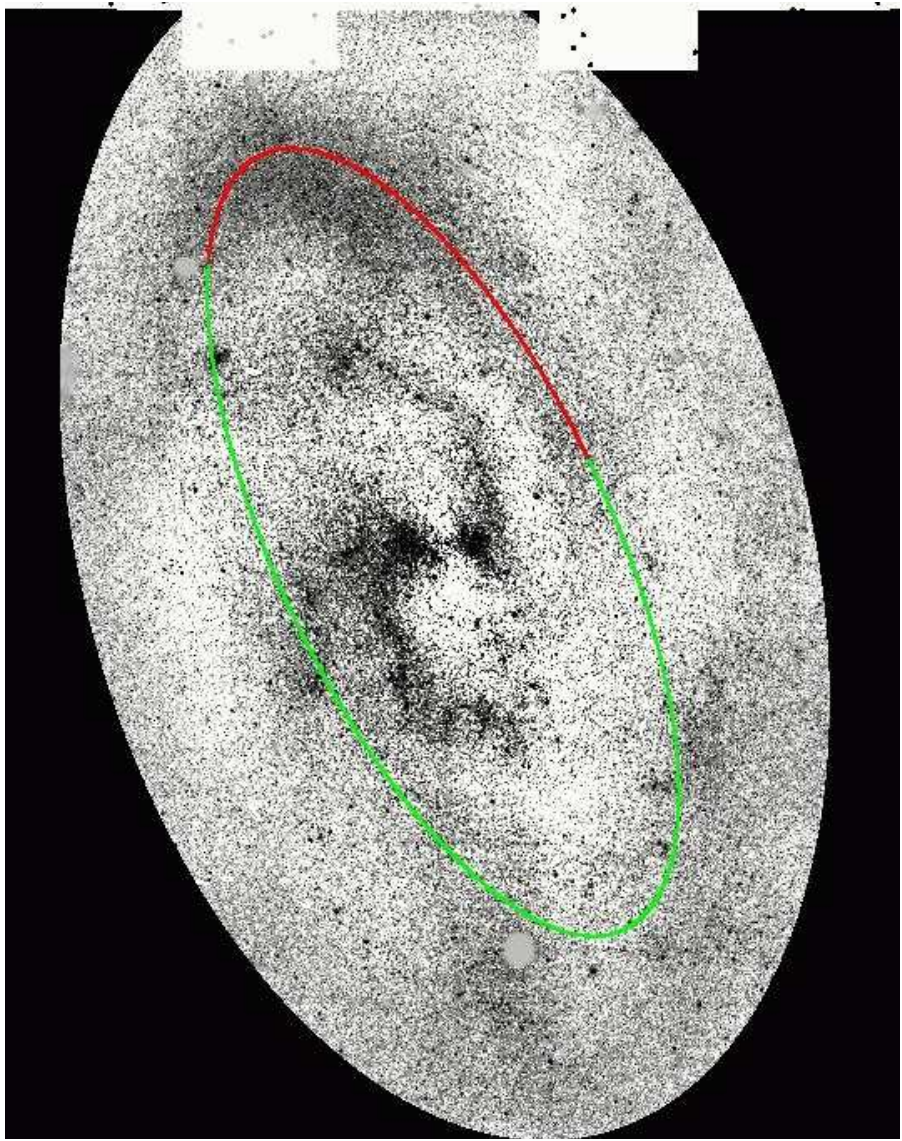}
\caption{A partial ring of very red stars is seen in this J+H+K$_{\rm s}$ image
of M33, secured from a special set of 2MASS observations
wherein the integration time was increased by a factor of six. The
ring is seen to full advantage with a simple ellipsoid model
subtracted. Black areas are those of excess emission while white areas correspond to areas of deficit
emission.
The northern plume or arc spans up to $5'$ in width and is
located at a radius of 2 to 3 disk scale lengths (or 12 to 18 arcmin).
An ellipse (of semi-major axis 16 arcmin) is drawn to pass through the arcs;
 North is up and and East to the left.}
\label{m33_2mass}
\end{figure*}

In an earlier study (Block et al. \cite{blocketal04}),
we presented  near-infrared images of M33 from a deep subsample of 
2MASS, in which individual stars were unresolved. The deep 2MASS 
images revealed remarkable arcs or spiral arms of red stars in the
outer disk; the northern arc subtends $120^\circ$ in azimuth angle
and $\sim 5'$ in width (see Figure \ref{m33_2mass}). The northern
arc is dominant although a very faint southern counterpart arc, forming
a partial ring, can also be seen. The arcs lie at a radius of 2-3 disk
\hbox{scale} lengths (in V, the disk scale length is 6 arcmin; Ferguson et al.
\cite{fergusonetal06}). 
Surprisingly, Fourier analysis of the deep 2MASS images 
showed that the dominant $m=2$ peak did not correspond to any inner 
spiral arm morphology seen in optical photographs, for which M33 is
so famous. Rather, the $m=2$ peak corresponds to the giant outer red arms in the
2MASS images (see Figure \ref{m33_2mass}). 


Block et al. (\cite{blocketal04}) noted that the very red colour of the arcs
could not be due to extinction by dust grains.
Freedman, Wilson and Madore (\cite{freedmanetal91}) derive a mean value for the total colour excess 
for the Cepheids in M33 of E(B-V) = 0.10 $\pm$ 0.09 mag, which
includes both foreground (Milky Way) and internal M33 extinction.  The
K$_{\rm s}$-band extinction A(K$_{\rm s}$) is approximately one-tenth that in the
optical (Rieke \& Lebofsky \cite{rieke85}), so the dust extinction at K$_{\rm s}$ is only 
$\sim$ 0.03
mag. Extinction by dust grains cannot 
generate the very red colours found in the arcs in the outer disk of M33, where the extinction at
K$_{\rm s}$ would be even smaller.

Our methodology differs from conventional ways of
photometrically identifying carbon stars using intermediate band
filters (V, I, 77, 81) as presented by Rowe et al. (\cite{roweetal05})
for M33, using the Canada-France-Hawaii Telescope (CFHT). 
With such a method, one needs a large enough telescope to
resolve individual stars. 
The (V, I, 77, 81) photometric method was pioneered by Cook, Aaronson 
and Norris (\cite{cooketal86}); another viable set of filters to identify carbon
stars are the Sloan filters. Demers \& Battinelli 
(\cite{demers05}) used the three Sloan filters 
{\it g, r, i}, but again resolution of stars is crucial.
Demers and Battinelli (2005) also used the  CFHT
with the Megacam camera to identify 361 
new C-star candidates in M31.

Some investigators (e.g. Tsalmantza et al. \cite{tsalm06})  have focussed their
attention on identifying carbon stars on the basis of the {\it red
J-K$_{\rm s}$ tail} (see Figures 1, 4, 6 and 9 in Tsalmantza
et. al. 2006). The usefulness of such a methodology is that near-infrared all 
sky surveys secured with moderate sized telescopes can potentially be exploited
to identify carbon stars in spiral galaxies well outside our Local Group, where
individual stars are not resolved.   

The arcs seen in Figure 1 are identified on the basis of their red
J$-$K$_{\rm s}$ colours. They lie around 15 arcmin in radius,
between 12 and 18 arcmin, or 2-3 disk scale lengths, whereas the actual
disk truncates at 5 scale lengths (Ferguson  et al. \cite{fergusonetal06}).
They are therefore like spiral arms in the outer disk,
as seen also by Rowe et al. (\cite{roweetal05}). 
The latter authors have identified the C and M-type stars in
the AGB population, and traced the Carbon star to M star ratio (C/M) as a function of radius.
This ratio is not only an index of age but also of metallicity, increasing in 
metal-poor regions. Indeed, a C-star requires that the O-dominated surface
be reversed to C-dominated, due to the dredged-up material from
the star interior. This is easier to do when the surface is metal-poor.
Rowe et al. (\cite{roweetal05}) show that the C/M ratio increases with radius
and then flattens beyond 20 arcmin (or 5 kpc), indicating a metallicity
gradient covering most of the disk, including the outer spiral arcs.

Our JHK$_{\rm s}$ photometric method uses the fact that the colour of the 
northern arc extended to very red colours of \hbox{J$-$K$_{\rm s}$$~> 1.1$}.
It was argued that while very old M
giants of solar abundance can indeed reach \hbox{J$-$K$_{\rm s}$ $\sim$ 1} (see e.g. Figure 2
in Bessell and Brett \cite{besselbrett98}) and even redder if they are
super-metal-rich (see Frogel and Whitford \cite{frogelwhitford87}), stars with \hbox{J$-$K$_{\rm s}$ $> 1$} in
low-metallicity regions cannot be M-giants but
rather, very red carbon stars. There is a strong radial metallicity
gradient in M33 (-0.16 dex/kpc in O/H, over 4-5kpc, see Beaulieu et
al. \cite{beaulieu06}). Beaulieu et al. have
confirmed this metallicity gradient, already found
in HII regions, B-supergiants or planetary nebulae, by detecting beat Cepheids in
M33. An excellent discussion of the metallicity gradient in M33 may be found in Magrini, Corbelli 
and Galli (\cite{magrini07}).
The outer regions of M33 are relatively metal-poor, and solar abundance is reached
only in the very central domain of M33. 

In this paper, we examine the robustness of our 
photometric technique as a stepping stone to exploring
the outer disks of more distant spiral galaxies,
wherein individual carbon stars may be
present, but unresolved. To this end, we need follow-up
spectroscopy. Here we discuss exploratory Keck I and Keck II 
spectroscopic observations of a few of the
reddest and intrinsically brightest stars in the northern arc. All
seven of our target candidates have very red \hbox{J$-$K$_{\rm s}$} colours and
should be confirmed to be TP-AGB carbon stars if our methodology is correct. 

\begin{figure*}[ht]
\vspace{15.0cm}
\includegraphics{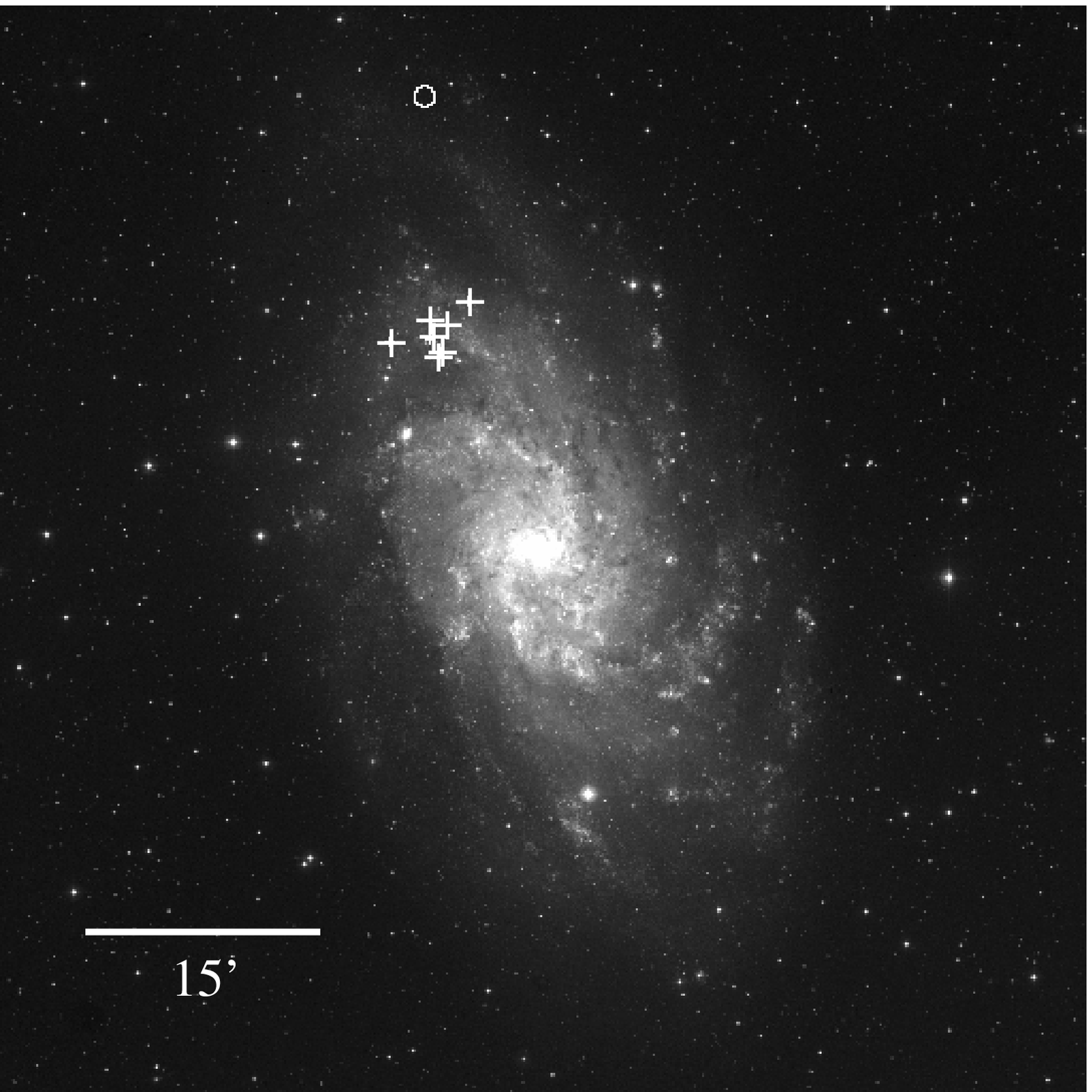}
\caption{The positions of the seven stars listed in Table \ref{table_mag_cstars} are indicated
  by plus signs; the stars lie in the northern swath of carbon bearing
  stars identified photometrically in an earlier study (Block et
  al. \cite{blocketal04}). 
 Spectra of these stars (see Fig. \ref{spectra_7stars})
were secured using the Keck I telescope in Hawaii. Also observed
  with its twin sister telescope, Keck II, are two stars 
   which lie 0.5 degrees away from the centre of M33 in an area
  identified by a white circle. These outlying stars are representative of a
  very red family identified in near-infrared imaging with the Hale
  5m reflector at Mount Palomar. In this DSS image, North is up and East is to the left.}
\label{m33_plus_carbonstars}
\end{figure*}

\section{Observations}
As a prelude to conducting the Keck spectroscopy, we imaged a
section of the northern  plume of M33 
with the Hale 5m reflector at Mount Palomar. We used the 2048$\times$2048 
array near-infrared camera  WIRC (Wilson et al. \cite{wilsonetal03}), 
to resolve individual stars. 
The field of view is $8.5'\times8.5'$ with  0.25 arcsec pixels.  
The seeing FWHM is 0.8$''$ in J, and 0.7 $''$ in 
the K$_{\rm s}$ band.  The telescope was centered at 
01h34m28.1s,  +30d54m00s (J2000). 
The total $JHK_{\rm s}$ integration time was $\sim$ 9 minutes, 
 reaching a limiting surface brightness at K$_{\rm s}$ of 23.2 
mag per pixel$^2$,  or 20.2 mag per arcsec$^2$ 
(4$\times$4 pixels gives 1 square  arsec).
The point source photometry  $S/N=10$ limits are 19.0, 18.0, 16.9 mag in JHK$_{\rm s}$, respectively. 

Table \ref{table_mag_cstars} lists the magnitudes and colours of 7 candidate targets in the outer northern
arc seen in Block et al. (\cite{blocketal04}). Their positions are indicated by plus 
signs (+) in Figure \ref{m33_plus_carbonstars}. In deriving the absolute magnitudes in Table \ref{table_mag_cstars},
we assume a distance modulus to M33 of
24.64$^{m}$, corresponding to a linear distance of 840 kpc (Freedman
et al. \cite{freedmanetal91}). Noting that \hbox{E(B$-$V) = 0.10 $\pm$ 0.09 mag}
(Freedman et al. \cite{freedmanetal91}), and that the K$_{\rm s}$-band extinction is approximately
one-tenth that in the optical (Rieke and Lebofsky \cite{rieke85}), we use a 
dust extinction correction for the K$_{\rm s}$ apparent magnitudes of 0.03 mag.    

\begin{figure*}[ht]
\vspace{20.0cm}
\includegraphics{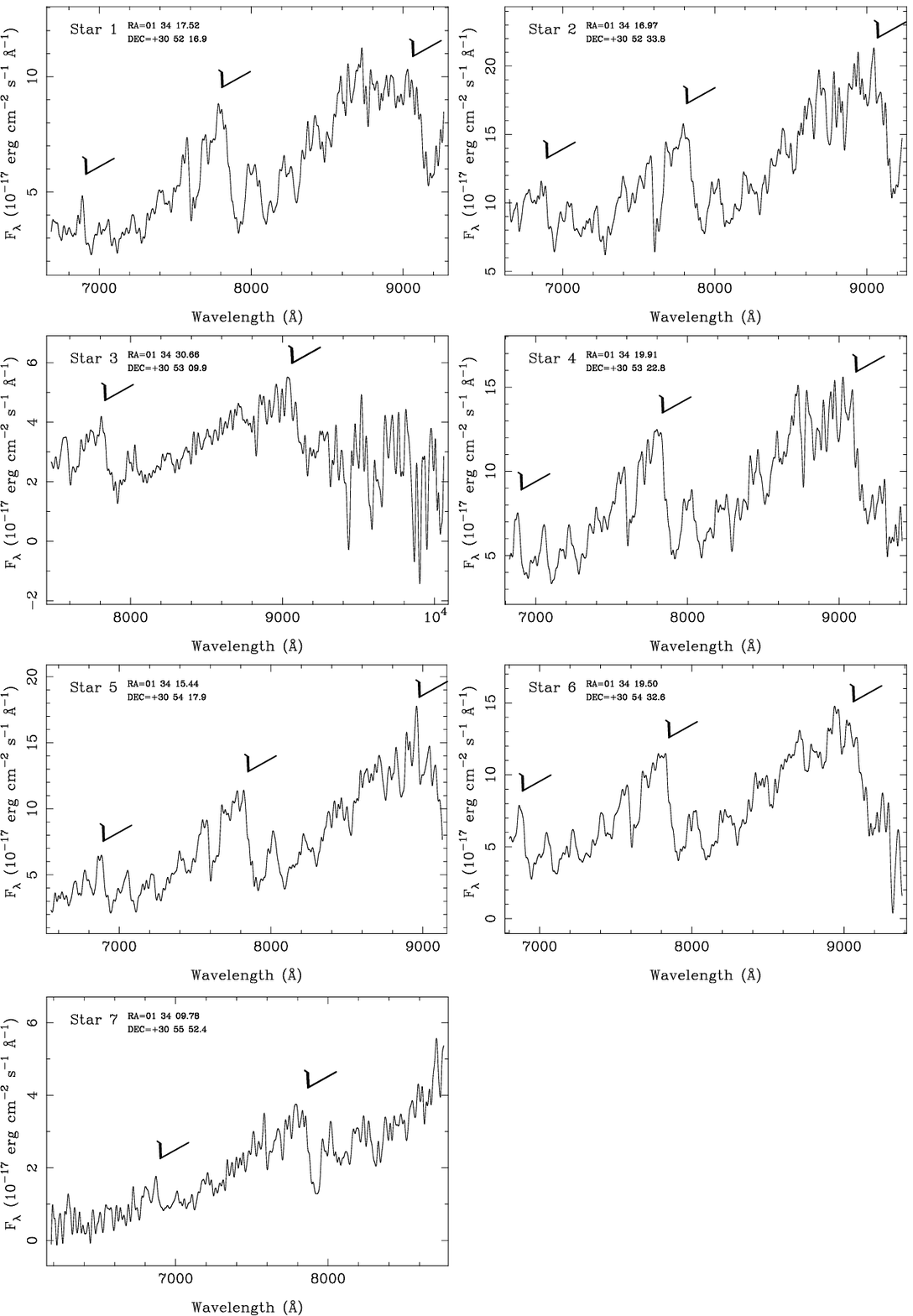}
\caption{Spectra of the seven stars seen in Figure \ref{m33_plus_carbonstars},
secured using the Keck I telescope. 
All C stars candidates are very red, with \hbox{J$-$K$_{\rm s}$$>$2 mag};
their individual colors are listed in Table \ref{table_mag_cstars}.
The coordinates of each star are indicated in each
  panel. Tick marks identify the prominent CN bands found in each
  star, shortward of 7000\AA\ and 8000\AA\ and longward of
  9000\AA . Each star is spectroscopically confirmed to belong to the 
  thermally-pulsing asymptotic giant branch carbon star family.}
\label{spectra_7stars}
\end{figure*}

\begin{table}
  \caption[]{Magnitudes and colors of our target stars in the northern red arc of M33}
   \begin{center}
 \begin{tabular}{cccccccc}
\hline
Star & J  &  H &     K$_{\rm s}$   &   J-K$_{\rm s}$  &  J-H   & H-K$_{\rm s}$ & M$_{\rm K_{s}}$ \\
\hline
1 & 17.96 & 16.76 & 15.91 &  2.05  & 1.20  & 0.85 & -8.74 \\
2 & 18.49 & 17.09 & 16.32 &  2.16  & 1.40  & 0.77 & -8.33 \\
3 & 19.61 & 18.04 & 16.66 &  2.96  & 1.57  & 1.39 & -7.99 \\
4 & 18.57 & 17.26 & 16.49 &  2.08  & 1.31  & 0.77 & -8.16 \\
5 & 18.82 & 17.35 & 16.54 &  2.28  & 1.47  & 0.81 & -8.11 \\
6 & 18.42 & 17.24 & 16.46 &  1.97  & 1.18  & 0.78 & -8.19 \\
7 & 18.37 & 16.91 & 15.82 &  2.55  & 1.46  & 1.10 & -8.83 \\
\hline
 \end{tabular}
\end{center}
\label{table_mag_cstars}
\end{table}

Spectra of these carbon star candidates were obtained on 2004 Aug 17
UT with the Low Resolution Imaging Spectrograph (LRIS; Oke et al.
\cite{okeetal95}) attached to the Keck I telescope. The aperture mask slitlets had widths and lengths
corresponding to 1\farcs2 and 11\farcs5, respectively. The 600 groove
mm$^{-1}$ grating gave a resolution of $\sim8$ \AA\ per projected slit
width and a spectral range from 6825 to 9415 \AA\ for slitlets near
the centre of the mask.  The stars were dithered along the slitlets
between the 2 exposures, each 600 seconds long.  The spectrum from each
slitlet was extracted and reduced independently, the wavelength
calibration being derived from airglow lines. An approximate flux
calibration was produced by observing the spectrophotometric standard
star G191B2B (Massey et al. \cite{masseyetal88}) in one of the slitlets near the
centre of the mask and transfering the system response function found
for this slitlet to the same wavelength intervals for the other
slitlets. This approach neglects any differential vignetting or other
field-dependent variations in response, but these are expected to be
small for LRIS in any case.

Also secured with the Keck II telescope were spectroscopic observations of an area 0.5 degree away from the centre
of M33 (circled in Figure \ref{m33_plus_carbonstars}). Near-infrared imaging with the
Hale 5m reflector at Palomar had revealed stars with very red \hbox{J$-$K$_{\rm s}$} colours
in this outer low metallicity region. The data were obtained on 2004 July 18 UT using the
Echellette Spectrograph and Imager (ESI; Sheinis et al. \cite{sheinisetal02}) on the
Keck II telescope.  We used the echellete mode which allows a wavelength
coverage from 3900 to 11060 \AA\ and yields a dispersion of
$\sim$ 56 km s$^{-1}$ with the 1$''$ slit.  The spectra were obtained at
parallactic angle near transit, and the total exposure was 1200 seconds  
for each object.

The spectra were reduced using a combination of J. Prochaska's IDL  
reduction package, ESIRedux, and the echelle tasks in IRAF.
The spectra were flux calibrated using spectrophotometric standards from
Massey et al. (\cite{masseyetal88}) also observed with the slit at the parallactic  
angle.

Observations of M33 were also made using all four bands of the Infrared
Array Camera (IRAC) mounted on the Spitzer Space Telescope (Werner et al.
\cite{werner04}; Gehrz et al. \cite{gehrz07}) on 2005 January 21 as part of a
Spitzer Guaranteed Time Observing Program (Program ID 5) conducted by
Spitzer Science Working Group member R. D. Gehrz.

The IRAC instrument (Fazio et al. \cite{fazioetal04}) is composed of four detectors
that operate at 3.6$\mu m$ (channel 1), 4.5$\mu m$ (channel 2), 5.8$\mu m$ (channel 3) and 8.0$\mu m$ (channel 4).
  All four detector arrays
are 256$\times$256 pixels in size with mean pixel scales of 1.221, 1.213, 1.222,
and 1.220$'' \rm{pixel}^{-1}$ respectively.  The IRAC filter band centers are at
3.548, 4.492, 5.661 and 7.87 $\mu m$ and the adopted zero magnitude fluxes
are 280.9 Jy (channel 1), 179.7 Jy (channel 2), 115.0 Jy (channel 3) and
64.1 Jy (channel 4) as described in the IRAC Data Handbook 
V3.0 (see Gehrz et al. 2007).

The M33 mapping sequence consisted of 438 frames per channel, including a
3 point 1/2 pixel dither for each map position. The integration time
was 12 seconds per frame.

The raw Spitzer data were processed and flux calibrated with
version 11.0.2 of the Spitzer Science Center (SSC) pipeline.  Post-BCD
processing was carried out using an artifact mitigation algorithm
developed by Carey (\cite{carey05}) and
the 2005 September 30 Linux version of the
SSC MOPEX software (Makovoz and Khan \cite{makovozkhan05}). The artifact mitigation
algorithm alleviates the effects of muxbleed, column pulldown/pullup,
electronic banding, and bias variations between the images. Three
additional corrections were implemented with MOPEX:
Background Matching, Outlier Detection, and Mosaicing.
Background matching was performed by minimizing the pixel differences in
overlapping areas with respect to a constant offset computed by the
program. Cosmic rays and other outliers were detected and eliminated
with the Outlier Detection module.
In the final step the images were
reinterpolated to a pixel scale of approximately 1.224$'' \rm{pixel}^{-1}$  and
mosaiced to create a final image spanning approximately 1.2$^{\circ}$ x 1.4$^{\circ}$.

\section{Analysis of the Keck I and Keck II spectra}

Even for nearby M33, identification of carbon stars by spectroscopic means
necessitates the use of the largest of groundbased telescopes. 
The first carbon star to be
spectroscopically confirmed in M33 dates back to the work of Mould and
Aaronson (\cite{mouldaaronson86}; Figure 7) who used the Hale 5m reflector at
Palomar. The 
individual V
magnitudes of each target star in Table \ref{table_mag_cstars} are estimated, from their
JHK$_{\rm s}$ colours, to be fainter 
than 22$^{m}$, with one star (star 3 in Figure \ref{spectra_7stars})
as faint as V$\sim$ 25$^{m}$.
The spectra sample only the bright end of the K$_{\rm s}$-band luminosity function
for carbon stars, which ranges from M$_{\rm K}$ $\sim$ -6 to -9.
Much longer integration times would be needed for intrinsically fainter
carbon stars.

Carbon stars show a
plethora of molecular spectral features, including the C$_{2}$ Swan 
bands and the CN
bands. The presence
of the hugely dominant CN bands shortward of 7000\AA\  and 8000\AA\
and longward of 9000\AA\ are indicated by tick marks in Figure
\ref{spectra_7stars} and unambiguously reveal the C-star status for each of these seven
stars.


\section{Analysis of the ISM dust morphology in M33}

\begin{figure}
\vspace{12.0cm}
\includegraphics{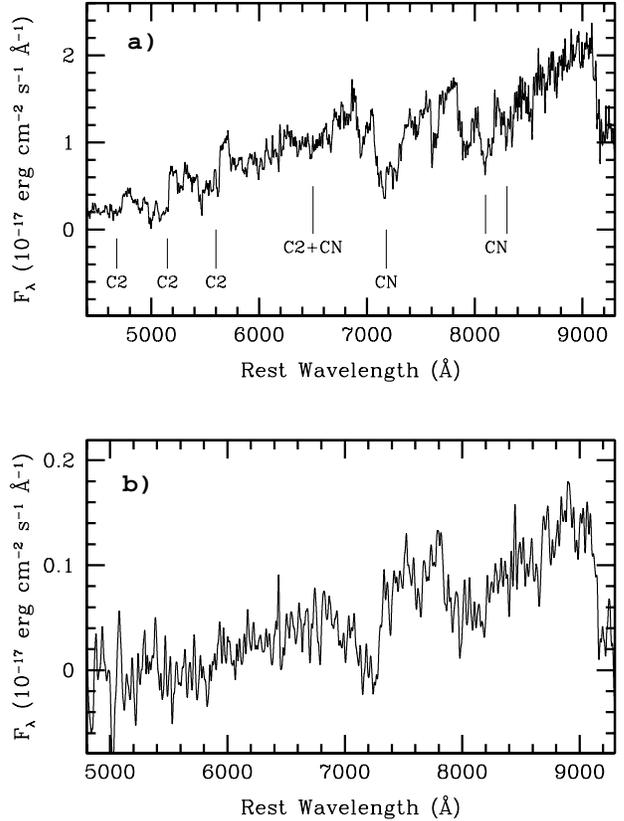}
\caption{Spectra with the Keck II telescope, of 
two stars 0.5 degree ($\sim$5 scalelengths) away from the centre
  of M33 (the field circled in Figure \ref{m33_plus_carbonstars}), centered at approximately
  RA(2000)=1h34m21s and DEC(2000)=31$^{\circ}$08$'$45$\farcs$7. The dominant CN
  absorption features are clearly evident and betray the signature of
  TP-AGB carbon stars. }
\label{spectra_binary}
\end{figure}

\begin{figure}
\vspace{12.0cm}
\includegraphics{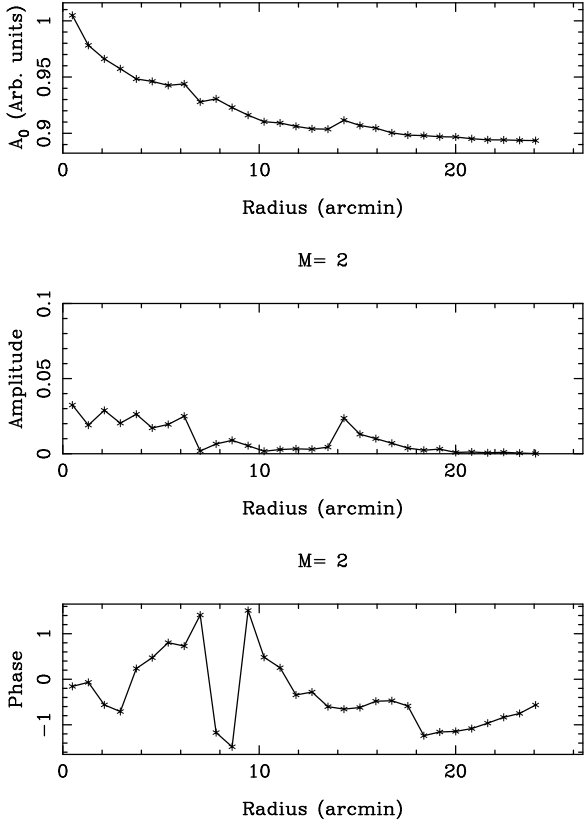}
\caption{The topmost panel shows a dust emissivity 8$\mu m$ radial
  profile. In the middle panel is shown the amplitude of the Fourier
  m=2 mode as a function of radius. The local peak at $r=14'$
  coincides with the inner radial limit of the arcs. 
The outer spiral arms or ``arcs" correspond to the higher m=2
amplitude between 14 and 18 arcmin. The phase of the
  m=2 component is presented in the lower panel. We see a change of
  phase in the dust emissivity distribution at 18 arcmin, at
the outer boundary of the outer spirals, precisely 
 the same radius where Corbelli and Schneider
  (\cite{corbellischneider97}) find a warp in HI. }
\label{m33_ampli_phase}
\end{figure}

The carbon star arcs are clearly seen in JHK$_{\rm s}$ images, but the
global ISM dust morphology can also be
effectively traced at 8$\mu m$. The reason is as follows:
as carbon stars pulsate, their atmospheres become extended
and matter may reach a temperature low enough for some
elements to condense into very small particles (``star dust"). 
An outflow develops as a result of
radiation pressure and the star is progressively
surrounded by an expanding circumstellar shell of dust and gas
(Le Bertre et al. \cite{lebertreetal03}). Emission from the shells
of both carbon stars and also dust from O-rich shells is detected at
8$\mu m$ (Buchanan et al. \cite{buchanan06}). Buchanan and co-workers note that ``most of the C-rich stars have spectra that are
dominated by warm dust" and these authors show that stars such as MSXLMC587 (O-rich) and MSXLMC1488 (C-rich)
both show emission in the IRAC 8 $\mu m$ window (see their Figure 2, for example). The trend is 
for the flux to climb from longer ($\sim$ 38$\mu m$) to the shorter
wavelength of 8$\mu m$. As an interesting aside, one does not need high ultraviolet excitation to produce 
infrared emission features:
stars as cool as 3000 K can generate
PAH features, for example, as discussed by Li and Draine (\cite{draine02}).

Carbon stars are one of the most 
important contributors to the replenishment of dust in the 
interstellar medium. Our IRAC image of M33 at wavelength 8$\mu m$ offers
unprecendented insight into the global ISM dust morphology.  
Deprojection of  our IRAC images of M33 were facilitated by adopting log R$_{25}$ = 0.23(RC3) 
where R is the ratio of the major to minor axis, and a position angle of 23 degrees 
(Deul and vd Hulst \cite{deul87}, Regan and Vogel \cite{regan94}). 
In order to remove the small contribution of starlight
(at the end of the Rayleigh-Jeans tail)
from the 8 $\mu m$ image, we subtract a scaled 
version of the IRAC 3.6 $\mu m$ map.
As in Block et al. (\cite{blocketal06}), we generate a `non-stellar' pure dust
map from
8$\mu m$ - $\kappa$ 3.6$\mu m$, where $\kappa$ (a constant) assumes the
value of 0.25. This subtraction removes the contribution from stellar
photospheres and leaves only the emission from dust grains, which trace
the ISM morphology. The
major uncertainties entering into the precise value of $\kappa$  are, firstly, the extended
aperture correction,
and secondly, the exact mid-IR colour of the stellar populations which one 
is modeling by
scaling the 3.6 $\mu m$  emission. The second correction 
introduces uncertainties of order 0.05 mag or perhaps larger (Ashby, private communication). 
A slightly smaller value of $\kappa$ (0.232 instead of 0.25) has been suggested by Helou and 
co-workers (\cite{helou04}),
but our use of a slightly higher photospheric contribution (at a level of 2 percent) is well within the
correction uncertainties; the choice
of $\kappa$ is only accurate at the 10-15 percent (and not one or two percent) level.

A complimentary method could, of course, be to focus our attention on
photospheric emission, and not ISM dust emission. At the shorter IRAC 
wavelength of 3.6$\mu m$, carbon star photospheres would
be detected, but so would the photospheres of O-rich stars. A full
presentation and analysis of all IRAC images of M33 will be presented
elsewhere (Gehrz et al., in preparation). In this paper, we confine our
attention to the dust grain morphology, as we did for the Andromeda
Spiral in Block et al. (\cite{blocketal06}).

The Fourier method has been extensively discussed in a number of papers (e.g.,
Consid\`ere and Athanassoula \cite{consat82}; 
Puerari and Dottori \cite{puedot92}, among others). In the Fourier method, an image is
decomposed into a basis of logarithmic spirals
of the form $r$=$r_o {\rm exp} (-{m\over p} \theta)$.
The Fourier coefficients $A(p,m)$ can be written as

$$ A(p,m) = \frac{1}{D}\int_{-\pi}^{+\pi}\int_{-\infty}^{+\infty} I(u,\theta)
{\rm exp}[-i(m \theta + p u)] d u d \theta$$

Here, $u \equiv {\rm ln}\;r$, $r$ and $\theta$ are the polar coordinates,
$m$ represents the number of the arms, $p$ is related to the pitch angle
$P$ of the spiral by  $P$=${\rm atan}(-m/p)$, and $I(u,\theta)$ is the
distribution of light of a given deprojected galaxy, in a
\hbox{(${\rm ln}\;r$, $\theta$)} plane. $D$ is a normalization factor
written as

$$ D=\int_{-\pi}^{+\pi} \int_{-\infty}^{+\infty} I (u, \theta) d u d \theta $$

In practice, the integrals in $u \equiv {\rm ln}\;r$ are 
calculated from a minimum radius (selected to exclude the bulge where there
is no information of the arms) to a maximum radius (which extends to the outer
limits of the arms in our images).

The inverse Fourier transform can be written as

$$S(u,\theta) = \sum_m S_m (u) {\rm e}^{im \theta} $$

\noindent where

$$ S_m(u) = \frac{D}{{\rm e}^{2u} 4 \pi^2} \int_{p_-}^
{p_+}\;G_m (p) A(p,m) {\rm e}^{i p u}\;dp$$

\noindent and $G_m(p)$ is a high frequency filter used to smooth
the $A(p,m)$ spectra at the interval ends (see Puerari and Dottori
\cite{puedot92}), and it has the form

$$ G_m(p) = {\rm exp} \left[ -\frac{1}{2} \left( \frac{p - p_{max}^m}{25}
\right)^2 \right] $$

\noindent where $p_{max}^m$ is the value of $p$ for which the amplitude of
the Fourier coefficients for a given $m$ is maximum. The chosen interval ends
($p_+$=$+50$ and $p_-$=$-50$), as well as the step-size $dp$=$0.25$, are
suitable for the analysis of galactic spiral arms.

Firstly, the Fourier $m=2$ mode in the 8$\mu m$ image also does not
correspond to the inner spiral arm morphology, but rather shows a local
peak at a radius of 14$'$, just longward of 2 scalelengths. This is
precisely where the outer red arcs were detected by Block et
al. (\cite{blocketal04}).

Another intriguing fact is that there is a {\it distinct change of
phase} in the ISM dust morphology at
3 disk scale-lengths (18$'$), which is precisely the region of the
outer boundaries of the ``arcs" or outer spiral arms, where Corbelli and
Schneider (\cite{corbellischneider97}) find the prominent HI-warp.   

We believe it fully consistent to propose that the very red,
and relatively metal-poor stars have recently been formed by gas
infall which is inextricably tied to its strong HI-warp. 
It seems highly plausible that 
fresh low-metallicity gas is being fed to the host
galaxy M33. The outer disk from which mass is accreted
is inclined to the inner disk of M33, and has a different angular
momentum, as revealed by the tilted disk model.

Our conclusions are compatible with the work of Rowe et al.
(\cite{roweetal05}) who
published a spatial map of the C/M-star ratio, that reveals conspicuous
maxima at exactly the positions of
what we call ``arcs", or outer spiral arms. They interpret the radial
distribution of the C/M ratio
uniquely as a metallicity indicator, and conclude that there is a strong
metallicity gradient up to 5kpc (20 arcmin)
and then the gradient flattens out. Their interpretation is in terms of
the viscous disk model, where the outer shear of the
rotation curve produces mixing in the outer parts, able to wash out
abundance gradients.

The viscous disk model certainly applies to M33 to some extent, and is
certainly a needed ingredient of the
external gas accretion model. The flattening of the metallicity gradient
could be linked
to the accretion of low-metallicity gas.
The viscous accretion disk model has been frequently invoked  to build
exponential galaxy disks, either with viscous torques
to exchange angular momentum, or with gravitational torques due to
spiral arms or bars. In the latter case the equivalent
viscosity is called "gravitational viscosity" (e.g. Lin \& Pringle
\cite{linpringle87}).
 If the viscous and star formation time scales are of the same order,
then the resulting stellar disk has an exponential distribution
independent of the disk rotation law and of the assumed viscosity
prescription.
The expected metallicity gradient is also exponential, as computed by
Tsujimoto et al. (\cite{tsuji95}). As for M33, it is highly possible that
the ``viscous disk" model
applies until a radius of 5kpc (20 arcmin), i.e. just outside
the arcs, and is the mechanism with which the accreted gas coming from
the warped
component is consumed into stars and disk formation. The optical disk is
exponential up to this radius.


\section{Conclusion}

We have confirmed, through Keck spectroscopy of individual stars,
the presence of an enhanced abundance of carbon stars in the 
outer spiral arms in M33, in particular in the two ``arcs" identified
previously from their red colour (Block et al. 2004).
 Our conclusions are in very good agreement with the recent work of
Rowe et al. (2005), who have produced a spatial map of the 
C/M-star ratio in M33 showing the conspicuous outer arms.
We also see the outer spiral arms in the Fourier m=2 component of the Spitzer
 8$\mu m$ dust emission image. These arms
correspond to the radius where the HI-21cm warp starts, and we propose 
an interpretation in terms of recent star formation, fueled from the 
gas accreted from this outer reservoir. A very similar conclusion, in terms of M33 accreting 
gas, has recently
been reached by Magrini, Corbelli and Galli (\cite{magrini07}), on the basis of O/H, S/H 
and [Fe/H] abundances
in the Triangulum Spiral Galaxy M33.

The HI gas reservoirs in the outer parts of galaxies  
are observed outside nearly all spiral disks (Sancisi
\cite{sancisi83}).  Their conspicuous warped morphology,
even in the absence of any perturbation or companion,
implies that this gas is being almost continuously accreted,
with a different angular momentum than that of the inner disk 
(e.g. Binney 1992).
Keres et al. (\cite{keresetal05}) discuss cold gas
accretion along filaments in the cosmic web. Accreting systems
in gas need not show any signs of accretion in stars, such as the presence
of tidal tails, stellar loops or close companions. 

The implications of the presence of carbon stars in the outer disk of M33
immediately beckons the question of a possible ubiquity of such an
intermediate age population in the outer domains of other spiral disks
out to the Virgo cluster and beyond. 

At hand is the potential to 
apply our photometric technique of identifying thermally-pulsing
asymptotic giant branch carbon stars to more distant spirals on the basis
of their red J-K$_{\rm s}$ colours in low metallicity domains, an approach
adopted by other teams such as Tsalmantza et al. (\cite{tsalm06}). 
It is a sobering thought that it takes a groundbased 
class 8$-$10m telescope to secure  
individual carbon star spectra in the outer disk of M33, our second closest
spiral, in realistic amounts of observing time. Hence the urgent need
to identify unresolved sets of carbon stars in distant spiral
galaxies, on the basis of colours alone - without resort to 
resolving such stars first with thirty metre class telescopes, 
and then imaging them through Sloan or (V, I, 77, 81) filters.

Carbon stars may make an important contribution to the IR luminosity
of high-redshift galaxies.  Using the Large Magellanic Cloud as a
guide, carbon stars are produced in large numbers between ages of
about 0.6 and 2 Gyr.  Therefore galaxies that undergo a burst of
star formation will have their IR light boosted 0.6 Gyr later, and
this extra light will die away after about 2 Gyr. We know that star
formation in the early universe 
(Bouwens et al. \cite{bouwensetal03}) was already proceeding
at redshift of at least \hbox{z $\sim 6$}.  There may be an epoch starting about 0.6
Gyr after the onset of star formation in the universe (i.e. at
redshifts \hbox{z $\sim 4$}) characterized by large numbers of carbon
stars. The dominant output from the carbon stars will be redshifted
into the mid-infrared where these stars could double the observed
extragalactic flux. Maraston (\cite{maraston05}) indeed proposes the use of 
carbon stars as an age indicator for high-redshift stellar populations. 
Instruments like the Mid-Infrared Instrument on
board the JWST are needed to image such galaxies in their (rest-frame)
2.2 micron band.

\acknowledgements{This paper was completed whilst DLB was a Visiting Professor at the 
 Mount Stromlo and Siding Spring Observatories, Canberra; the hospitality of Ken and 
Margaret Freeman is very
warmly acknowledged.
DLB and IP are indebted to the Anglo American Chairman's Fund, 
Mr. C. Sunter, Mrs. M. Keeton and the Board of Trustees.
IP acknowledges support from the Mexican foundation CONACyT
under project 35947--E.
This paper is based in part on data obtained at the W. M. Keck Observatory, which is
operated as a scientific partnership among the California Institute of
Technology, the University of California and the National Aeronautics
and Space Administration. The Observatory was made possible by the
financial support of the W. M. Keck Foundation.
 This work is also partially based on observations made with
the Spitzer Space Telescope, which is operated by the Jet Propulsion
Laboratory, California Institute of Technology under a contract with
NASA. RDG and CEW were supported by NASA through an award
issued by JPL/Caltech. DLB warmly thanks the Vice-Chancellor of the University of the Witwatersrand
for the Vice-Chancellor's  Research Award in 2006. DLB also expresses much gratitude to Fani Titi of the TISO Foundation 
for his stellar support. We thank   
M. Ashby, P. Wood and S.P. Willner for their input.
Finally, we thank the 
anonymous referee for insightful comments.


\begin{thebibliography}{ }

\bibitem[2006]{beaulieu06}
Beaulieu J-P., Buchler J., Marquette J-B. et al: 2006, ApJ 653, L101
 
\bibitem[1998]{besselbrett98}
Bessell, M.S., Brett, J.M. 1998, PASP 100, 1134

\bibitem[1992]{binney92}
Binney, J.: 1992,  ARAA, 30, 51


\bibitem[2004]{blocketal04}
Block, D.L., Freeman, K.C., Jarrett, T.H., Puerari, I., Worthey, G., Combes, F., Groess, R. 2004,
A\&A 425, L37

\bibitem[2006]{blocketal06}
Block, D. L., Bournaud, F., Combes, F., Groess, R., Barmby, P., Ashby, M. L. N., 
Fazio, G. G., Pahre, M. A., Willner, S. P.: 2006, Nature 443, 832

\bibitem[2003]{bouwensetal03}
Bouwens, R.J., Illingworth, G.D., Rosati, P. et al. 2003,
ApJ 595, 589
 

\bibitem[2006]{buchanan06}
Buchanan, C.L., Kastner, J.H., Forresst, W.J., Hrivnak, B.J., Sahai, R., Egan, M., Frank, A.,
Barnbaum, C. 2006, AJ, 132, 1890


\bibitem[2005]{carey05}
Carey, S. 2005, {\tt http://spider.ipac.caltech.edu/
staff/carey/irac\underline{ }artifacts/}

\bibitem[1982]{consat82}
Considere, S., Athanassoula, E.: 1982, A\&A 111, 28

\bibitem[1986]{cooketal86}
Cook, K.H., Aaronson, M., Norris, J. 1986,
ApJ 305, 634

\bibitem[1997]{corbellischneider97}
Corbelli, E., Schneider, S.E. 1997,
ApJ 479, 244

\bibitem[2005]{demers05}
Demers, S., Battinelli, P.: 2005, A\&A 436, 91

\bibitem[1987]{deul87}
Duel, E.R., van der Hulst, J.M.: 1987, A\&A Supp, 67, 509



\bibitem[2004]{fazioetal04}
Fazio, G.G., Hora, J.L., Allen, L.E. et al. 2004,
ApJS 154, 10

\bibitem[2006]{fergusonetal06}
Ferguson, A., Irwin, M., Chapman, S., Ibata, R., Lewis, G., Tanvir, N. 2006,
astro-ph/0601121

\bibitem[1991]{freedmanetal91}
Freedman, W.L., Wilson, C.D., Madore, B.F. 1991, ApJ 372, 455

\bibitem[1987]{frogelwhitford87}
Frogel, J.A., Whitford, A.E. 1987, ApJ 320, 199

\bibitem[2007]{gehrz07}
Gehrz, R. D. et al. 2007, Rev. Sci. Instrum. 78, 011302


\bibitem[2004]{helou04}
Helou, G. et al. 2004, ApJS, 154, 253


\bibitem[2005]{keresetal05}
Keres, D., Katz, N., Weinberg, D.H., Dav\'e, R. 2005
MNRAS 363, 2

\bibitem[2003]{lebertreetal03}
Le Bertre, T., Tanaka, M., Yamamura, I., Murakami, H. 2003
A\&A 403, 943


\bibitem[2002]{draine02}
Li, A., Draine, B.T. 2002, ApJ 572, 232



\bibitem[1987]{linpringle87}
Lin, D. N. C., Pringle, J. E.: 1987,  MNRAS 225, 607



\bibitem[2007]{magrini07}

Magrini, L., Corbelli, E., Galli, D.: 2007, A \& A (in press) (astroph/07043187)



\bibitem[2005]{makovozkhan05}
Makavoz, D., Khan, I. 2005, in ``Mosaicing with MOPEX'' Astronomical Data
Analysis Software and Systems XIV, ASP Conference Series,
Eds. P. L. Shopbell, M. C. Britton, and R. Ebert, {\sl in press}

\bibitem[2005]{maraston05}
Maraston, C. 2005, in Multiwavelength mapping of galaxy formation and evolution,
Eds. A. Renzini and R. Bender, Springer, Berlin, 290

\bibitem[1998]{masseyetal88}
Massey, P., Strobel, K., Barnes, J.V., Anderson, E. 1988,
ApJ 328, 315

\bibitem[1986]{mouldaaronson86}
Mould, J., Aaronson, M. 1986, ApJ 303, 10

\bibitem[1995]{okeetal95}
Oke, J.B., Cohen, J.G., Carr, M., Cromer, J., Dingizian, A.,
Harris, F.H., Labrecque, S., Lucinio, R., Schaal, W., Epps, H., Miller, J. 1995,
PASP 107, 375

\bibitem[1992]{puedot92}
Puerari, I., Dottori, H. A.: 1992, A\&AS,  93, 469

\bibitem[1994]{regan94}
Regan, M.W., Vogel, S.N.: 1994, ApJ, 434, 536


\bibitem[1985]{rieke85}
Rieke, G.H., Lebofsky, M.J. 1985, ApJ 288, 618

\bibitem[1976]{rog76}
Rogstad, D. H., Wright, M. C. H., Lockhart, I. A.: 1976,
ApJ 204, 703

\bibitem[2005]{roweetal05}
Rowe, J.F., Richer, H.B., Brewer, J.P., Crabtree, D.R. 2005,
AJ 129, 729

\bibitem[1983]{sancisi83}
Sancisi, R. 1983, Internal Kinematics and Dynamics of Galaxies, IAU Symp. 100, 55

\bibitem[2002]{sheinisetal02}
Sheinis, A.I., Bolte, M., Epps, H.W., Kibrick, R.I., Miller, J.S., Radovan, M.V., Bigelow, B.C., Sutin, B.M. 2002,
PASP 114, 851

\bibitem[2006]{tsalm06}
Tsalmantza, P., Kontizas, E., Cambr\'esy, L., Genova, F., Dapergolas, A., Kontizas, M.: 2006,
A\&A 447, 89

\bibitem[1995]{tsuji95}
Tsujimoto, T., Yoshii, Y., Nomoto, K., Shigeyama, T.: 1995, A\&A 302, 704

\bibitem[2004]{werner04}
Werner M.W. et al. 2004, ApJS 154, 1

\bibitem[1991]{wilson91}
Wilson, C.D. 1991, AJ 101, 1663

\bibitem[2003]{wilsonetal03}
Wilson, J.C., Eikenberry, S.S., Henderson, C.P. et al. 2003,
Proceedings of the SPIE 4841, 451

\end{thebibliography}
\end{document}